\documentclass[conference]{IEEEtran}
\ifCLASSINFOpdf
   \usepackage[pdftex]{graphicx}
   \graphicspath{{../pdf/}{../jpeg/}{figures/}}
   \DeclareGraphicsExtensions{.pdf,.jpeg,.png}
\else
   \usepackage[dvips]{graphicx}
   \graphicspath{{../eps/}}
\fi
\usepackage[cmex10]{amsmath}
\usepackage{url}

\usepackage[utf8]{inputenc}
\usepackage{listings}
\usepackage{float}
\usepackage{color}
\usepackage{xcolor}
\usepackage{textcomp}
\usepackage{multirow}
\usepackage{verbatim}
\usepackage{amssymb}
\usepackage{dsfont}
\makeatletter
\def\lst@makecaption{%
    \def\@captype{figure}%
  \@makecaption
}
\makeatother

\colorlet{punct}{red!60!black}
\definecolor{background}{HTML}{EEEEEE}
\definecolor{delim}{RGB}{20,105,176}
\colorlet{numb}{magenta!60!black}

\lstdefinelanguage{json}{
    basicstyle=\normalfont\ttfamily,
    numbers=left,
    numberstyle=\scriptsize,
    stepnumber=1,
    numbersep=8pt,
    showstringspaces=false,
    breaklines=true,
    frame=lines,
    backgroundcolor=\color{background},
    literate=
     *{0}{{{\color{numb}0}}}{1}
      {1}{{{\color{numb}1}}}{1}
      {2}{{{\color{numb}2}}}{1}
      {3}{{{\color{numb}3}}}{1}
      {4}{{{\color{numb}4}}}{1}
      {5}{{{\color{numb}5}}}{1}
      {6}{{{\color{numb}6}}}{1}
      {7}{{{\color{numb}7}}}{1}
      {8}{{{\color{numb}8}}}{1}
      {9}{{{\color{numb}9}}}{1}
      {:}{{{\color{punct}{:}}}}{1}
      {,}{{{\color{punct}{,}}}}{1}
      {\{}{{{\color{delim}{\{}}}}{1}
      {\}}{{{\color{delim}{\}}}}}{1}
      {[}{{{\color{delim}{[}}}}{1}
      {]}{{{\color{delim}{]}}}}{1},
}

\begin{document}
%
\title{Dynamic Allocation of Serverless Functions\\ in IoT Environments}

\author{
\IEEEauthorblockN{Duarte Pinto}
\IEEEauthorblockA{
DEI - Faculty of Engineering,\\ University of Porto\\ Porto, Portugal\\ Email:
pinto.duarte@fe.up.pt}
\and
\IEEEauthorblockN{João Pedro Dias}
\IEEEauthorblockA{INESC TEC and \\
DEI - Faculty of Engineering,\\ University of Porto\\ Porto, Portugal\\ Email: jpmdias@fe.up.pt}
\and
\IEEEauthorblockN{Hugo Sereno Ferreira}
\IEEEauthorblockA{INESC TEC and \\
DEI - Faculty of Engineering,\\ University of Porto\\ Porto, Portugal\\ Email: hugosf@fe.up.pt}}


%


\maketitle

\begin{abstract}
The IoT area has grown significantly in the last few years and is expected to
reach a gigantic amount of 50 billion devices by 2020\nocite{kn:Cisco01}. The
appearance of serverless architectures, specifically highlighting \textit{FaaS},
raises the question of the suitability of using them in IoT environments. Combining IoT with a serverless architectural design can be effective when trying to make use of the
local processing power that exists in a local network of IoT devices and creating a
fog layer that leverages computational capabilities that are closer to the
end-user. In this approach, which is placed between the device and the
serverless function, when a device requests for the execution of a serverless function will decide based on previous metrics of execution if the serverless function should be executed locally, in the fog layer of a local network of IoT
devices, or if it should be executed remotely, in one of the available cloud servers. Therefore, this approach allows dynamically allocating functions to the most suitable layer.

\end{abstract}



%
\IEEEpeerreviewmaketitle

\section{Introduction}
Continuing the current trend, mobile data usage is expected to keep increasing
exponentially, part of it due to mobile video streaming and
IoT. The estimation is that the number of data that was generated
by mobile devices during the year of 2017 exceeded the $\displaystyle 6 * 10^9$ Gb
per month. Together with the traffic generated by laptops and peer-to-peer
communications, overall traffic demand might reach $\displaystyle 11 * 10^9$ Gb per
month\cite{kn:Dehos2014} \cite{kn:Baresi2017}. To compute such a big amount of
data, cloud computing would appear to be the obvious solution but there are cases
where the latency that comes with transmitting data back and forth might be
a limitation. In certain situations, it is also not feasible to expect a constant and
reliable internet connection to an always-on server, either because it might not
be economically wise or because it might not be infrastructurally possible. In
order to solve the need for low latency, as well as to improve fault tolerance, by
not relying on an always-on centralized server, serverless architectures and fog
computing aim to reduce the dependency on the cloud by making more use of the
local resources of a device and improving communication between local devices,
only leaving the data-intensive tasks to the cloud\cite{kn:Baresi2017}. With the
increasing trend in serverless solutions, such as AWS Lambda, it is opportunistic
to implement these concepts in IoT.

Although IoT has been around for a few years already, the same cannot be said
about services that provide cloud solutions and cloud infrastructure for rent. 
Likewise, when Serverless and Fog Computing solutions first appeared their
usefulness and benefits for the IoT ecosystem was obvious and developers began to
mix them together in order to get the most out of this new trend.

Despite its success and the promising future for the mix of this concepts, the
area is still fairly new and few solutions can take advantage of the processing
power in the cloud and in the local network of IoT devices in an efficient way
without compromising speed. It is already possible to have a network of IoT
devices working together to execute a series of serverless functions, but not all
serverless functions are suitable to run on low-end devices. To choose where each
serverless functions should be executed (locally or in the cloud) is a manual task
and the end result is that developers choose to have all serverless functions
running in the cloud as it is easier to manage and less risky. Nonetheless, there
is a lot of potential processing power dormant in each local network that could
be used to improve response times, improve fault-tolerance,  and to slash costs of
hosting cloud processing infrastructure. 

\section{Background}

\subsection{Internet of Things}\label{sec:dialecto}
The \textit{Internet of Things} is a term given to the network of ever increasing
number of mundane objects with embedded systems, that allows them to interact with
each other or with someone remotely. This network creates a smarter and
self-regulated environment that depends less on the input of a physical entity and
more on the input of other \textit{things}, so that it removes unnecessary steps
and it frees the user from dealing with mundane tasks.

\subsection{Serverless}
Defining the term serverless can be difficult, as the term is both misleading and
its definition overlaps other concepts, such as Platform-as-a-Service (PaaS) and
Software-as-a-Service(SaaS). Serverless stands in between these two concepts, where
the developer loses some control over the cloud infrastructure but maintains
control over the application code \cite{kn:Baldini}.

"The term ‘Serverless’ is confusing since with such applications there are both
server hardware and server processes running somewhere, but the difference to
normal approaches is that the organization building and supporting a ‘Serverless’ application is not looking after the hardware or the processes - they are outsourcing this to a vendor." 

\hfill Mike Roberts

\hfill 2016

The most important area of serverless, for this paper, is
\textbf{Function-as-a-Service (FaaS)} in which, the server-side logic is still
written and controlled by the developers, but they run in stateless containers
that are triggered by events, ephemeral and are fully managed by the 3rd party
entity. Despite being a recent paradigm, there is already some investigation about
the performance and usability of serverless solutions managed by 3rd party
entities \cite{kn:Lee}.

\subsection{Fog Computing}
Fog Computing is a virtual resource paradigm, located in between the Cloud
layer(traditional cloud or data centers) and the Edge layer (smart end devices) in
order to provide computing power, storage, and networking services. Although
conceptually located in between the two layers, in practice, this platform is located at the
edge of the network \cite{kn:Bonomi}. "This paradigm supports vertically-isolated,
latency-sensitive applications by providing ubiquitous, scalable, layered,
federated, and distributed computing, storage, and network connectivity"~\cite{kn:Iorga2017}.

\section{State of the Art}

\subsection{Exploration vs Exploitation and Multi-Armed Bandit} \label{sec:sota_mab}
Exploration vs Exploitation is a common decision making dilemma, both in real life
and in the computer world. Choosing between a known good option or taking the risk
of trying an unknown option in the hope of finding a best result is a choice that
we try to balance in the hope of minimizing the total regret(total opportunity
lossi\cite{kn:Silver}) we face.

If we had access to all the information about the universe in question, we could
either brute-force or use other smart approaches to achieve the best results. In
this situation, the problem comes from only having \textit{incomplete}
information. In order to make the best overall decisions, we need to
simultaneously gather enough about the system and keep the total regret at a
minimum. Exploitation will choose the best known option in order to avoid any
regret. Exploration will take the risk of choosing one of the less explored
options with the purpose of gathering more information about the universe in
question, reducing short-term success for long-term success. A good strategy will
use both options, exploration and exploitation, to achieve the best results.

This problem is similar to the one approached in this paper, where there is no
knowledge about the environment and we need to mix both exploration and
exploitation to gather the best results and to achieve the best possible result.

The Multi-Armed Bandit is a known problem that exemplifies the Exploration vs
Exploitation dilemma. The problem places us with multiple slot machines, each with
a different reward probability. Given the setting, the objective is to find
\textit{the best strategy to achieve the highest long-term reward}\cite{kn:Weng2018}.

The goal is to maximize the total reward, $\sum^T_{t=1}r_t$, or in other words,
minimize the regret of not taking the optimal action in every step.

The optimal reward probability $\theta^*$ of the optimal action $a^*$ is:
\begin{displaymath}
    \theta^* = Q(a^*) = \max_{a \in A}  Q(a) = \max_{1 \leq i \leq K} \theta_i
\end{displaymath}

\subsubsection{$\epsilon$-Greedy}
This algorithm will choose the best known action most of the times but it will
also explore randomly from time to time. The value of an action is given by \cite{kn:Weng2018}:

\begin{displaymath}
    \hat{Q}_t(a) = \frac{1}{N_t(a)} \sum_{\tau=1}^t r_\tau \mathds{1}[a_\tau=a]
\end{displaymath}

$\mathds{1}$ is a binary indicator function and $N_t(a)$ represents the total number of
times that a given action as been selected, $N_t(a)= \sum_{\tau=1}^t
\mathds{1}[a_\tau=a]$. 

In this algorithm, we take the best known action most of the times, $\hat{a}^*_t
= \arg \max_{a\in A}\hat{Q}(a)$, or , with a probability of $\epsilon$, we take a
random action. The best known action will be taken with a probability of
$1-\epsilon$.

\subsubsection{Upper Confidence Bounds}
Random exploration might not be the best option because it might lead to trying an
action that was previously concluded as bad. One way of avoiding is is to give
priority to options with a \textit{high degree of uncertainty}, actions for which
there isn't a confident value estimation yet.

Upper Confidence Bounds (UCB) will translate this potential into a value, the
upper confidence bound of the reward value, $\hat{U}_t(a)$ \cite{kn:Weng2018}. The
true will be below $Q(a) \leq \hat{Q}_t(a) + \hat{U}_t(a)$. $\hat{U}_t(a)$ will
vary with the number of trials and a larger sample of trials will result in a
smaller $\hat{U}_t(a)$. With the UCB algorithm the next action to take in the will be:

\begin{displaymath}
a_t^{UCB} = \arg\max_{a\in A}\hat{Q}(a) + \hat{U}(a)
\end{displaymath}

To estimate the upper confidence bound, if prior knowledge of how the distribution
looks like can be discarded, then it is possible to apply \textit{"Hoeffding's
Inequality"}, a theorem that can be applied to any bounded distribution \cite{kn:Silver}.


Applying Hoeffding's Inequality to the rewards of the bandit will result in:

\begin{displaymath}
    U_t(a) = \sqrt{\frac{-\log p}{2N_t(a)}}
\end{displaymath}

\subsubsection*{UCB1}
To ensure that the optimal action is taken as $t\rightarrow\infty$, $p$ can be
reduced with each trial \cite{kn:Silver}.

In UCB1 algorithm replaces $p=t^{-4}$ resulting in:

\begin{displaymath}
    U_t(a) = \sqrt{\frac{2\log t}{N_t(a)}}
\end{displaymath}

The resulting algorithm, UCB1, is as follows

\begin{displaymath}
a_t = \arg\max_{a\in A}Q(a) + U(a) = \arg\max_{a\in A}Q(a) + \sqrt{\frac{2\log t}{N_t(a)}}
\end{displaymath}

\subsubsection*{Bayesian UCB}
\label{sota:bayesian_ucb}
The previous method, UCB1, does not make any assumptions about the reward
distribution $R$, only relying on Hoeffding's Inequality to make an estimation.
Knowing the distribution would allow or better estimates.

In the Bayesian UCB it is assumed that the reward distribution is Gaussian,
$R_a(r) = N(r;\mu_a,\sigma_a^2)$. The action that will give the best result is the
action that maximises standard deviation of $Q(a)$ \cite{kn:Silver}:

\begin{displaymath}
    a_t = \arg\max_{a\in A}\mu_a + \frac{c\sigma_a}{\sqrt{N(a)}}
\end{displaymath}

where the constant $c$ is an adjustable hyperparameter.

Bayesian UCB relies on a considerable ammount of prior knowledge of the rewards
and for it to be accurate \cite{kn:Silver}. Otherwise, it would be too
straightforward.

\section{Problem Statement}

As stated before in previous sections, not only is expected for the number of IoT
devices to grow immensely, both commercially and industrially, but there are
already many solutions that allow for serverless functions to be executed
remotely. Due to the nature of serverless functions, some of them could perfectly
be executed locally, using the joint processing power of the multiple IoT devices.
The hardship comes with using this power efficiently, having multiple serverless
functions, and knowing where to execute each one, locally or remotely. It is not
feasible for each developer to manually analyze performance across the different
runtime environments and make a decision where the function should be executed.
This is impeding the adoption of these concepts in IoT, despite the interest and
potential that exists in this evergrowing area. Not only there is a lack of
systems making use of serverless on-premises, the majority of the developers in
IoT opt for using the cloud for each and every need, disregarding the power that
exists locally.

Like what was presented above, there is a lack of practical know-how knowledge
available despite there being lots of incentives for it. There are lots of things,
but it is not easy to start developing a serverless IoT solution.

Given this, the aim of this project is to create an architecture for serverless IoT
platform and to build a proof of concept using existing open-source tools when
possible and avoiding proprietary solutions. The platform should:
\begin{itemize}
    \item Have a serverless cloud solution capable of answering HTTP requests from
        the \textit{things}.
    \item Make use of the local processing power of the multiple IoT devices to
        create a serverless virtual processing unit on the Fog Layer to answer local requests.
    \item Have multiple IoT devices with different functions capable of
        interacting with both the Cloud and Fog layer to execute different functions. 
\end{itemize}

\section{Proposed Solution}
\label{sec:proposed_solution}

\begin{figure}[ht]
  \centering
  \includegraphics[width=0.5\textwidth]{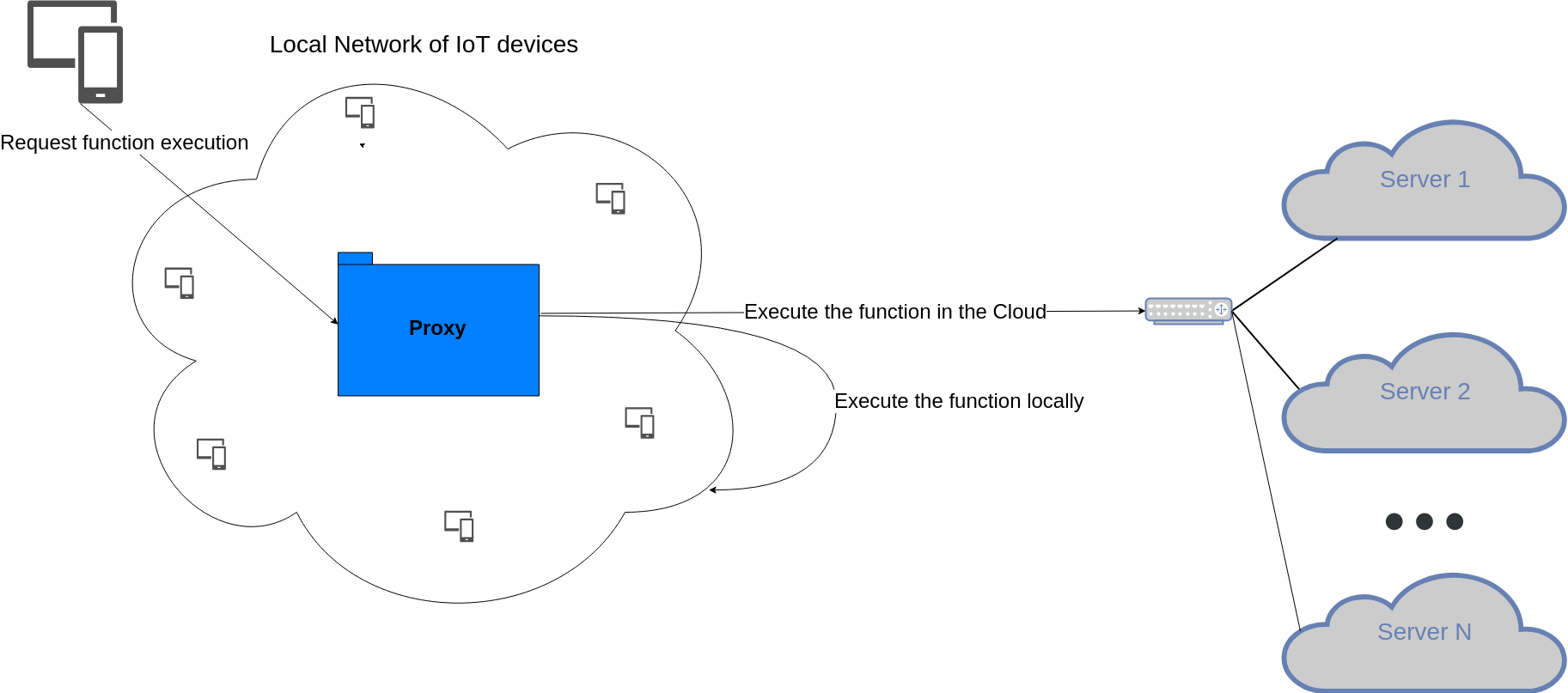}
  \caption{High level overview of project's architecture.}
  \label{fig:request_func_high_level_diagram}
\end{figure}

This project tries to fix the aforementioned problem by
introducing a proxy between the entity requesting the function execution and the
serverless function. This proxy will analyze each function's past history, by
looking at the time taken in past requests and make the decision of which runtime
environment \footnote{Runtime environment is the system where the serverless function will be executed, i.e., local network or one of the servers available in the cloud} should the request be forwarded to, see Figure
\ref{fig:request_func_high_level_diagram}. The proxy should be able to decide
between the local network of devices and one of the many available servers.

In order to improve fault-tolerance, in case of no Internet connection or if
one of the servers is not available if the request to the server fails, the proxy
should fallback to the local network. This way, even if the request to execute the function is forwarded to the server and fails, the function will still be executed locally.

The proxy is situated inside the local network of IoT devices and will forward the
request for a specific function to a gateway which forwards the function execution to
one of the IoT devices capable of executing the function. The load management,
containerization, replication, and clustering of the serverless functions is not
handled by the proxy, but it still has to be aware of the serverless functions
installed in the local network or any of the other runtime environments.

\subsection{Expected results and flow: Use cases}
\label{overview:usecases}
The following examples explain the expected results and decision-making of the
proposed solution. The decisions taken by the proxy are based only on previous
metrics of the time taken for the runtime environment to execute the function
(including network latency).

\subsubsection{Forward function execution to the cloud} \label{usecases:forward_cloud}

\begin{figure}[ht]
  \begin{center}
    \includegraphics[width=0.5\textwidth]{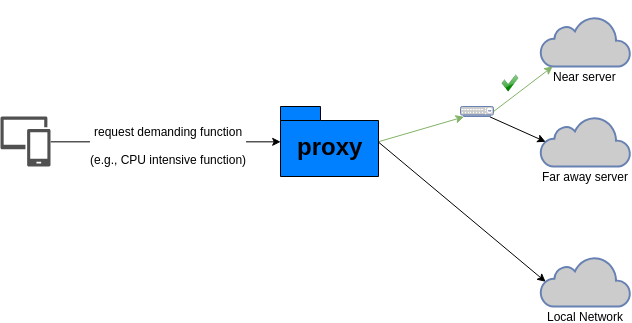}
    \caption{Request for the execution of a demanding function to be executed. The proxy will forward the request to the cloud because due to the high processing  power of the cloud server, the function will be executed more quickly. The nearest server was chosen because of latency.}
    \label{fig:succ-cloud-deploy}
  \end{center}
\end{figure}

The use case in Figure \ref{fig:succ-cloud-deploy} exemplifies a situation where
the requested serverless function is hardware intensive, therefore taking a lot of
time to execute locally. Due to the high processing power of the cloud servers, it
is beneficial to forward the execution request to one of the cloud servers, even
when considering the connection latency. From the multiple available servers, it
will opt for the one that is physically nearest (lower latency).

\subsubsection{Forward function execution to the local network}
\label{usecases:forward_local}

\begin{figure}[ht]
  \begin{center}
    \includegraphics[width=0.5\textwidth]{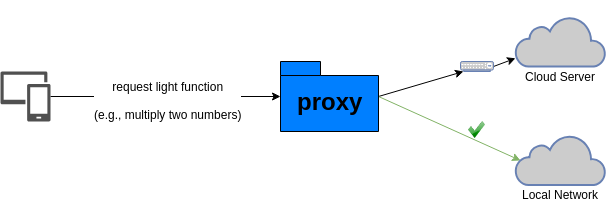}
    \caption{Request for the execution of a simple, light function to be executed. The proxy will forward the request to be run locally, as there is no benefit in
    executing the function on the cloud.}
    \label{fig:succ-local-deploy}
  \end{center}
\end{figure}

Contrary to the previous case, the Figure \ref{fig:succ-local-deploy} portrays a scenario where the requested serverless function is very light, being more beneficial to execute the function locally and avoid network latency. Despite the difference in power between the two environments, the previous metrics show that
the local environment is capable of satisfying the request more quickly.

\subsubsection{Fallback to the local network}
\label{usecases:fallback}

\begin{figure}[ht]
  \begin{center}
    \includegraphics[width=0.5\textwidth]{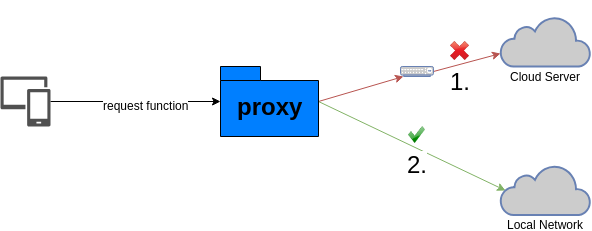}
    \caption{The request for the function execution to be in the cloud could not be satisfied (e.g., no internet connection). The proxy will then forward the
request for it to be executed in the local network.}
    \label{fig:fallback-local}
  \end{center}
\end{figure}

The Figure \ref{fig:fallback-local} depicts a scenario where the proxy first tries
to forward the request to one of the cloud servers (because it is more beneficial)
but fails in doing so. The proxy then decides to forward the request to the local
network, successfully completing the request. There are certain situations where
it is more favorable for the function to be executed on the cloud but it could
still be executed locally. Because it is not possible to always guarantee a
working connection, in these cases, if the connection fails the proxy will
fallback to execute function locally, assuring fail redundancy and the reliability
of the system.

\subsubsection{Manual forward. Bypass the weighting process}
\label{usecases:manual_forward}

It should also be possible for the developer to bypass the weighting process (the
evaluation of the different runtime environments) and manually choose where to forward the
request. This option should be possible either in the setup process of the
function or as an argument of the request for the function execution. 



\subsection{Weighting the runtime environments}

In order to make the decision of which runtime environment to forward the function
to, there has to be some weighting process that will weight each of the runtime
environments and compare them. This process will gather information about the
different runtime environments and then make an accurate estimation of which one
is the best choice (less time took). This is similar to the Exploration vs
Exploitation problem presented in \ref{sec:sota_mab}. Therefore, the following
algorithms were implemented to handle the weighting process:

\begin{itemize}
    \item \textbf{Greedy} - Has no exploration, simply assigns the weight as the
        mean average time taken.
    \item \textbf{UCB1} - Uses Hoeffding's Inequality to balance between
        exploration and exploitation, but the cumulative regret will still be considerable.
    \item \textbf{Bayesian UCB} - Analyzes the reward distribution to make a very
        accurate prediction of the weight but requires previous knowledge about the environments.
\end{itemize}

The developer or device can choose which of the weighting algorithms will be used
for that request, in the options, but the default algorithm is UCB1. UCB1 was
selected as default because despite Bayesian UCB being better, it requires
previous knowledge about the environment, as stated in \ref{sota:bayesian_ucb}.

\subsection{Code components}

\begin{figure}[ht]
  \begin{center}
    \includegraphics[width=0.5\textwidth]{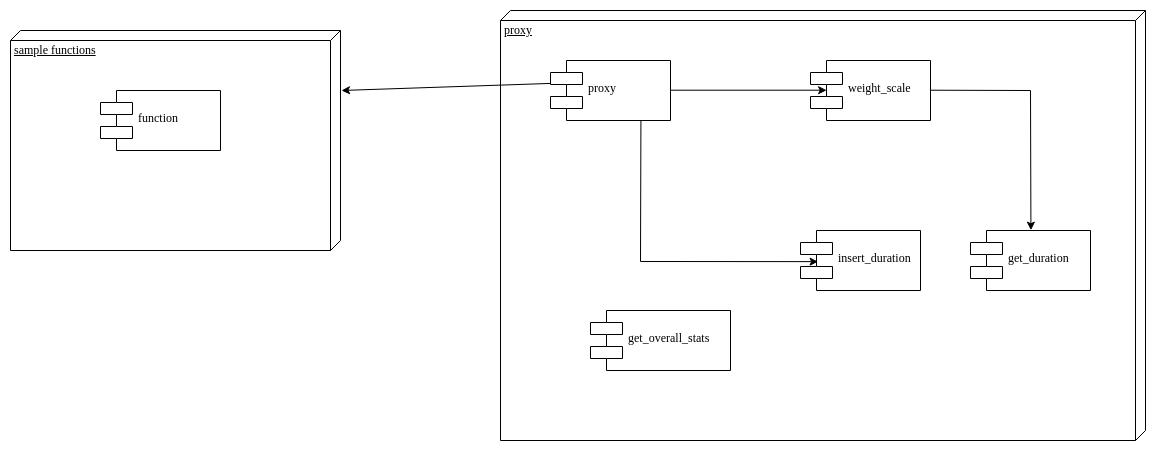}
    \caption{Component diagram of the project}
    \label{fig:component_diagram}
  \end{center}
\end{figure}

As it can be seen in Figure \ref{fig:component_diagram}, the proposed solution is
constituted of two main components, the \textbf{proxy} and the
\textbf{sample\_functions}. Each of these components is a package in itself.

\subsubsection{Packages}

\begin{itemize}
    \item \textbf{proxy} - The main package of the project responsible for
        all the logic. Deals with the reception of the request for the execution
        of a function, with the weighting of the environments in which the
        functions can run (locally or in the cloud in one of the multiple
        servers), and with the storage and retrieval of all metrics of previous
        function executions. 
    \item \textbf{sample\_functions} - The package that contains the serverless functions
        whose execution is going to be requested to the proxy package. This package is purely a sample with the purpose of simulating and analyzing resource demanding serverless functions (either light or heavy) and could be replaced by any other set of functions. The functions inside this package can be executed on either the local environment or on one of the servers (remote environments).
\end{itemize}

\subsubsection*{proxy}

All the different functions inside the proxy package communicate through HTTP and
expect to receive the content as application/json.

\begin{itemize}
    \item \textbf{proxy} - The main function through which all requests go through
        first. After receiving the list of weights associated with the execution of
        the requested function in each environment, it will choose the environment
        with the least weight and forward the execution of the serverless function
        to that runtime environment.
    \item \textbf{weight\_scale} - This function will analyze all the collected 
        metrics of the requested function and assign a weight to each runtime environment. It allows more than one algorithm for weight estimation.
    \item \textbf{get\_duration} - Retrieves the list of all the collected
        metrics of a function.
    \item \textbf{insert\_duration} - Store the time taken for a function to
        execute.
    \item \textbf{get\_overall\_stats} - Function that will return the summarized
        records of all the collected metrics for each function in each environment. Useful for analysis and evaluation of the results.
\end{itemize}

\subsubsection*{sample\_functions} \label{overview:sample_functions}

The serverless functions in this package have the purpose of simulating real serverless
functions for different purposes and execution times. The functions are aware of
the runtime environment they are being executed on and it is possible for them the answer
differently according to this. Here, the different time taken is simulated using a
\textit{wait} and using different values for different runtime environments.

\begin{itemize}
    \item \textbf{func\_light} - This function answers instantly, there should be
        no difference between executing the function locally or in the cloud,
        other than connection latency.
    \item \textbf{func\_heavy} - In this function there is a \textit{wait} of 2 seconds if
        it is executed locally or a \textit{wait} of 1 second if it is executed on the cloud. There should be no difference in the time taken across different cloud servers other than connection latency.
    \item \textbf{func\_super\_heavy} - similar to \textit{func\_heavy} but here the difference in time is bigger. There is a \textit{wait} of 4 seconds if the function is executed locally or a \textit{wait} of 2 seconds if the function is executed in the cloud.
    \item \textbf{func\_obese\_heavy} - This is a function that, due to its nature, it can
        only be executed in the cloud and its execution has been flagged as cloud-only. The proxy will not even try to run the function locally, it will always forward the request to the cloud. Because of this, there is no fallback to run locally.
\end{itemize}

\section{Experimentation and evaluation}
\subsection{Environment}
In this serie of experiments, the setup was configured with 3 runtimes environment.
1 local virtual machine running all the functions with Lubuntu 16.04, 2 cores and
1 Gb of RAM. There are also 2 cloud servers, one in London and other in Frankfurt,
both  hosted in a EC2 instance in AWS
\footnote{\url{https://aws.amazon.com/pt/ec2/instance-types/}}. They are both
t2.micro, running Ubuntu 16.04, with 1 core and 1 Gb of RAM.

\subsubsection*{Connection latency}
\label{res:conn_latency}
Both servers were pinged before the test to verify the connection latency. The
experiment is being made in Porto, Portugal. Here are the results: 
\begin{itemize}
    \item London - 71.153ms
    \item Frankfurt - 52.297ms
\end{itemize}

\subsection{First Experiment: With internet connection}
For this experiment both cloud servers were up and running and it were perfomed
99 iterations of requests. In each iteration was requested for every single one
of the serverless functions in \textit{sample\_functions} (see
section \ref{sec:proposed_solution} ) to be executed.
The system has no knowledge of the environment. The aim here is to identify
that it is accomplishing the mentioned use cases presented in sections \ref{usecases:forward_cloud},
\ref{usecases:forward_local}, and \ref{usecases:manual_forward}

It is expected that the proxy will forward requests for \textit{func\_light} to be
executed locally, for \textit{func\_heavy} requests to be executed either locally
or in the cloud. It will depend on the impact of the connection latency, but
generally, the connection latency should be less than 1 second (difference in time
that takes for the function to execute locally and remotely), which means that the expected result is for the proxy to choose to forward to one of the cloud servers.
\textit{func\_super\_heavy} is expected to be executed in the cloud most of the
times, due to the big difference in time taken, and the function
\textit{func\_obese\_heavy} should always be executed in the London server because
it is configured that way. Apart from \textit{func\_obese\_heavy}, some
exploration is expected for each of the different environments and not only
exploitation of the runtime environment that the system considers as the best
option. Because of the latency verified in \ref{res:conn_latency}, when choosing
between one of the servers, it should choose the Frankfurt server, because it is
the one with less latency (despite being physically further).

\subsubsection{Results}

The obtained results matched the expected results. For \textit{func\_light}, the
time taken was so small (less than 1/10 of a second) that proxy imediatly
converged in the best option. The results for the execution of the function
\textit{func\_light} translate the results expected for the use case
in section \ref{usecases:forward_local}.

For \textit{func\_heavy} and \textit{func\_super\_heavy}, it kept a ratio of
exploration vs exploitation of 3/7 and 5/6, respectively, but always choosing the
fastest of the cloud servers. The exploration rate also increases with the
duration of the execution, meaning that the proxy will look for better options the
longer it takes for a function to execute. The results for these two functions
correspond to the ones expected in the use case presented in section \ref{usecases:forward_cloud}.

Also, as expected, \textit{func\_obese\_heavy} had a 100\% accuracy, thus matching
the expected outcome stated in use case presented in section \ref{usecases:manual_forward}.

\subsection{Second Experiment: Without internet connection to the servers}
For this experiment, both servers were turned off and the internet connection was
cut, leaving the system only operational locally. The system still keeps all the
knowledge acquired in the previous experiment.
The aim here is to identify that it is accomplishing the use case
in section \ref{usecases:fallback}.

In this experiment, it is expected for the weighting algorithm to suggest
executing the serverless function in one of the cloud servers, because it will
lead to a faster execution. Because there is no internet connection, it is
expected for the proxy to try to execute the function remotely, fail, and then to
fall back to the local runtime environment. In the end, the function should be
executed in the local runtime environment leading to the request being answered
successfully.

\subsubsection{Results}
First, the function \textit{weight\_scale} is queried to know which of the runtime
environments the proxy is going to choose (because the proxy chooses the runtime environment with less weight, knowing the weights allow us to know which option
the proxy is going to take). Because there is more information about the system,
the weighting algorithm used was the Bayesian UCB. As it can be seen in the Listing
\ref{listing:exp_weight_query}, the runtime environment that is going to be
chosen is the Frankfurt's server.

\begin{lstlisting}[
    language=json,
    basicstyle=\footnotesize,
    caption= {Weights of the different servers, for the second experience, using the Bayesian UCB algorithm.},
    label=listing:exp_weight_query]
{
    "status": "success",
    "londonServer": 3.5747403999957266,
    "frankfurtServer": 1.1756422708191938,
    "local": 2.090245031544607
}
\end{lstlisting}

Even though the chosen runtime environment was Frankfurt's server, because there
was no internet connection it had to fall back to execute the function locally in
order to complete the request successfully, as seen in Listing
\ref{listing:exp_fallback_result}. The observed results match the ones expected
and also the proxy proceeded as stated in the use case presented in section \ref{usecases:fallback}.

\begin{lstlisting}[
    language=json,
    basicstyle=\footnotesize,
    caption= {The request was executed locally, as indicated by the key \textit{swarm},
which is the swarm (runtime environment) where the function was executed.
\textit{local}, is the name given to the local network of devices, as configured
when setting up the proxy.},
    label=listing:exp_fallback_result]
{
    "nodeInfo": "61e20a65b48e ",
    "swarm": "local",
    "message": "I was able to achieve this result 
    using HEAVY calculations",
    "status": "The light is ON"
}
\end{lstlisting}

\subsection{Third Experiment: Turning off Internet connection in a series of requests}
During this experiment, the main purpose is to run a cycle of requests and then
turn off the Internet access in the middle of the cycle to observe how this will
affect response times. It will be run 99 iterations of requests and in each
iteration, it will be requested the execution of \textit{func\_heavy}, After
request number \textit{50}, the Internet connection will be cut off, leaving the
system only operational locally. The system will keep all the knowledge gathered
in the previous experiments. The aim here is to identify that it is accomplishing
the use case in section \ref{usecases:fallback} and how results vary throughout.

In this experiment, it is expected for the weighting algorithm to suggest
executing the serverless function in one of the cloud servers, because it will
lead to a faster execution. Because of this, the mean total time of the request should
be smaller in the first 50 requests. After the 50th request, because the internet
connection was cut off, the proxy will try to execute the function remotely, fail,
then fallback to execute the function locally, resulting in a larger mean total
time. 

\subsubsection{Results}

\begin{figure}[h]
  \begin{center}
    \includegraphics[width=0.5\textwidth]{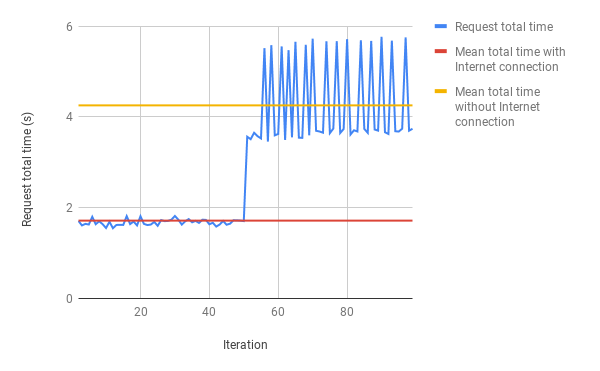}
    \caption{Chart illustrating the results of the third experience.}
    \label{fig:exp3_chart}
  \end{center}
\end{figure}

The obtained results, illustrated in Figure \ref{fig:exp3_chart}, match the
expected results. The mean total time of the request when there was Internet
connection was \textbf{1,71466602} seconds, and, at iteration number 50, it
jumped to \textbf{4,253785939} seconds when the Internet connection was cut off.
Despite having no internet connection, the system was still able to complete the
request, just with an added delay. The added delay was due to the fact that it had
to try to execute the function remotely and also because of the increased time it
takes to execute the function locally (2 seconds). 

\subsection{Fourth Experiment: Adapting to lag}
\label{res:exp4}
In this experiment, it is going to be executed a series of requests and after
reaching to a while, the Internet connection is going to be purposely slowed to
see how the system reacts in situations of lag and slow connection. It will be run
249 iterations of requests and in each iteration it will be requested the
execution of \textit{func\_heavy}, After request number \textit{50}, the Internet
connection will be slowed down (\textbf{28 kbps UP, 14 kbps DOWN}) and the system
will continue to be asked to execute the functions. The system will have none of
the knowledge gathered in the previous experiments. The aim here is to identify
that it is accomplishing the use cases in sections \ref{usecases:forward_cloud} and
\ref{usecases:forward_local}, and also that it is capable of adapting to changes
in the network.

In this experiment, it is expected that the system goes through three phases. In
the first phase, in the first 50 iterations, while the connection to the server is
working as expected, the system is supposed to gather information about the
environment and to converge to the best option (one of the cloud servers).

In the second phase, the Internet connection is slowed down and the execution of
function remotely should take longer than the execution of the function locally.
In this phase, the system is supposed to still converge to one of the cloud
servers but gradually diminishing the frequency in which it chooses the cloud
serves as the best option.

After this phase, the system will enter the third phase where the results gathered
after the introduction of network lag outweigh the results gathered in the first
phase. Here, the system should start to converge to the local network as the
best option.

\subsubsection{Final Remarks}

\begin{figure}[h]
  \begin{center}
    \includegraphics[width=0.5\textwidth]{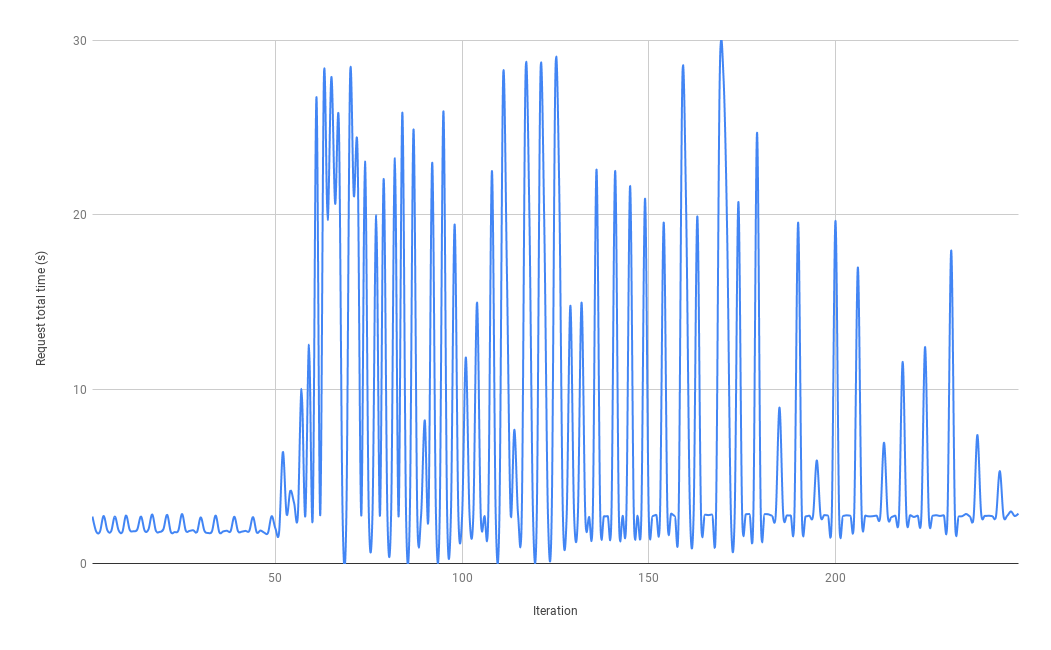}
    \caption{Requests total time throughout the various iterations of the
    fourth experiment.}
    \label{fig:exp4_total_time_chart}
  \end{center}
\end{figure}

The gathered results can be observed in Figure \ref{fig:exp4_total_time_chart}. As
expected, the results for the first 50 iterations are the expected results in a
standard situation. After the introduction of network lag, we start to observe
spikes in the total time it takes for the function to be executed. These spikes
refer to the execution of the function remotely. In the first 30/40 requests after
the introduction of network lag (iterations 50 to 90), the frequency of requests
that are executed remotely is still high. The frequency starts to diminish
from that point on and at around iteration 175 the system starts to choose the
local network more frequently than the cloud servers.

\begin{figure}[h]
  \begin{center}
    \includegraphics[width=0.5\textwidth]{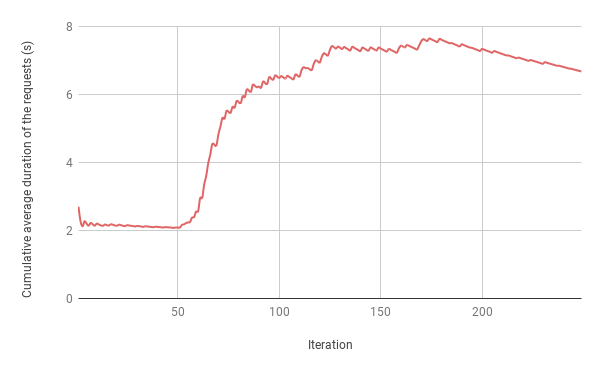}
    \caption{Cumulative average duration of the request throughout the various iterations of the fourth experiment. E.g., the value in iteration 50 (2.085604 seconds) is the average duration of the first 50 requests.}
    \label{fig:exp4_avg_chart}
  \end{center}
\end{figure}

Figure \ref{fig:exp4_avg_chart}, shows a different perspective of the results,
showing the cumulative average duration of the requests throughout the experiment.
It can be seen here that around iteration 175 the system changed course and
started to converge to the local network as the best option. This marks the point
where the system finally adapted to the changes introduced.

The various phases can be seen more easily here, in Figure
\ref{fig:exp4_avg_chart}. The first phase can be seen from iteration 0 to 50, the
second phase from iteration 50 to 175, and the third phase from iteration 175 to
250.

Nevertheless, it took around 125 iterations (from iteration 50 to iteration 175)
for the system to adapt. After 50 iterations where the system gathered information
that became invalid, it took the system 250\% more iterations to adapt to the new
conditions. These results show that the system, although capable of adapting, will
take a considerable amount of time to adapt to new conditions.

\section{Conclusion}
The experimentation and results presented in this paper go in accordance to
those expected and satisfy the proposed use cases in section \ref{overview:usecases}. The
developed solution is capable of analyzing the knowledge it has over the ecosystem
and will make a decision that will lead to a faster execution time and at the
same time explore different options that might lead to better results.
Additionally, the developed solution is also capable of detecting failures in the
remote execution of the serverless function and solve that problem, executing the
function locally and answering the request successfully.

In sum, the proposed solution proved to be capable of answering the demanded use
cases and the used approach was fruitful. Nonetheless, there are limitations and
some questions that this approach cannot answer.
Despite the fact that the proposed solution already reaches a level that is very
beneficial for most of the practical applications, further development must be
made to reach a more compelling solution.

Within the fields surrounding this work, there is a lot of uncharted
territory and unknown aspects. The choice between a serverless architecture and a
monolithic one is still not clear in all cases and adding these concepts and
infusing it with IoT and Fog Computing is a very new area. Because of this, the
principal contributions of this work were:

\begin{itemize}
    \item Innovative approach to the mix of IoT, fog computing and serverless. There is not much work in this mix of fields and this approach is both innovative
        and unseen.

    \item Enabling serverless both locally and remotely. The developer creating the serverless functions no longer has to actively choose where to deploy the functions, it is possible to automate that process and still keep total
        regret at a minimum.

    \item The ability to run serverless functions locally even if the connection to the server fails and improvement of fault-tolerance in systems.

    \item Gathering of existing knowledge, tools, and platforms suitable for developing solutions in the areas of IoT, serverless and fog computing.

    \item Development of a functional prototype built on top of widely used and mature solutions that can serve as inspiration for future and better solutions.
\end{itemize}

All the code and work made is openly available at \url{https://github.com/444Duarte/serverless-iot} .







\subsection{Further Work}

Although the work pursued so far, we consider that there is still a set of open research challenges that worth to be explored in the future. From those, the following are considered of more relevance:

\begin{itemize}
    \item \textit{Stateful}: Introducing statefulness and consistency across serverless
        different runtime environments is definitely a challenge but the end result would be of utmost usefulness and importance.
    \item \textit{Analysis of other metrics other than time} (e.g. energy consumption, CPU
        cycles, memory usage, network usage): The time taken is not the only metric that is important when choosing where to execute a serverless function. Analyzing the impact of other metrics would also be applicable in other different contexts and is something to take into consideration.
        A solution mixing different points of view that analyzed different metrics to choose the overall best result is also a possibility that would be of great use.
    \item \textit{Static analysis to verify the complexity of the functions}: A static analysis
        of the function could largely improve the exploration vs exploitation problem, allowing the system to start with some knowledge about the complexity of the function, diminishing the total regret.
\end{itemize}


\section*{Acknowledgment}

The  authors  acknowledge  the  financial  support  from  the Department  of
Informatics Engineering of the Faculty of Engineering  of  the  University  of
Porto  (FEUP/DEI)  and for  the  facilities  offered  to pursue this research
work.



\bibliographystyle{IEEEtran}
\bibliography{IEEEabrv,refs}
%



\end{document}